\documentclass[universe,article,accept,pdftex,moreauthors]{Definitions/mdpi} 
\newcommand{\rthis}[1]{\textcolor{black}{#1}}

\firstpage{1} 
\pubvolume{1}
\issuenum{1}
\articlenumber{0}
\pubyear{2025}
\copyrightyear{2024}
\datereceived{ } 
\daterevised{ } % Comment out if no revised date
\dateaccepted{ } 
\datepublished{ } 
\hreflink{https://doi.org/} % If needed use \linebreak
\Title{Constraints on Lorentz Invariance Violation from Gamma-Ray Burst Rest-Frame Spectral Lags Using Profile Likelihood}

\TitleCitation{Title}

\Author{Vyaas Ramakrishnan and Shantanu Desai*}

\AuthorNames{Firstname Lastname, Firstname Lastname and Firstname Lastname}

\isAPAStyle{%
       \AuthorCitation{Lastname, F., Lastname, F., \& Lastname, F.}
         }{%
        \isChicagoStyle{%
        \AuthorCitation{Lastname, Firstname, Firstname Lastname, and Firstname Lastname.}
        }{
        \AuthorCitation{Lastname, F.; Lastname, F.; Lastname, F.}
        }
}

\address[1]{Department of Physics, IIT Hyderabad, Kandi, Telangana-502284, India; vyaas3305@gmail.com (V.R.); shntn05@gmail.com (S.D.)}

\corres{Correspondence:  shntn05@gmail.com}

\abstract{We reanalyze the spectral lag data for 56 Gamma-Ray Bursts (GRBs)  in the cosmological rest frame to search for Lorentz Invariance Violation (LIV)  using frequentist inference. For this purpose, we use  the technique of  profile likelihood to deal with the  nuisance parameters, corresponding to a constant time lag in the GRB rest frame and an unknown intrinsic scatter, while the parameter of interest is the energy scale for LIV ($E_{QG}$).  With this method, we do not obtain a global minimum for $\chi^2$ as a function of $E_{QG}$ up to the Planck scale. \rthis{Thus}, we can obtain one-sided lower  limits on  $E_{QG}$ in a seamless manner. Therefore, the 95\% c.l. lower limits which we thus obtain on $E_{QG}$ are then  given by:  $E_{QG}\geq 2.07 \times 10^{14}$ GeV and $E_{QG}\geq  3.71\times 10^{5}$ GeV, for linear and quadratic LIV, respectively. }% Therefore, this work represents yet another  proof of principles application of profile likelihood in the search for LIV using GRB spectral lags.}

\keyword{Gamma-Ray bursts; Lorentz Invariance; Profile Likelihood } 

\begin{document}

%%%%%%%%%%%%%%%%%%%%%%%%%%%%%%%%%%%%%%%%%%
%\setcounter{section}{-1} %% Remove this when starting to work on the template.
%\section{How to Use this Template}

%The template details the sections that can be used in a manuscript. Note that the order and names of article sections may differ from the requirements of the journal (e.g., the positioning of the Materials and Methods section). Please check the instructions on the authors' page of the journal to verify the correct order and names. For any questions, please contact the editorial office of the journal or support@mdpi.com. For LaTeX-related questions please contact latex@mdpi.com.%\endnote{This is an endnote.} % To use endnotes, please un-comment \printendnotes below (before References). Only journal Laws uses \footnote.

% The order of the section titles is different for some journals. Please refer to the "Instructions for Authors” on the journal homepage.

\section{Introduction}

Spectral lags of gamma-ray bursts (GRB) have been widely used  as a probe of Lorentz Invariance Violation (LIV)~\citep{Desairev,WuGRBreview,WeiWu2}. The spectral lag is defined as the time difference  between the arrival of high energy and low energy photons, and is considered to be positive, if the high energy photons precede the low energy ones. In case of LIV caused by an energy-dependent slowing down of the speed of light, one expects a turnover in the spectral lag data at higher energies.

Most of the searches for LIV using GRB spectral lags have been carried out using fixed energy intervals in the observer frame. The first work to search for LIV using spectral lags between fixed rest frame energy bands was  the analysis in ~\citet{WeiWu} (W17, hereafter). This work considered a sample of 56 \rthis{Swift}-BAT detected GRBs, with spectral lags in the fixed rest frame energy bands: 100-150 keV and 200-250 keV~\cite{Bernardini}. Based on a Bayesian analysis, W17 obtained a robust lower limit on the LIV energy scale, $E_{QG} \geq 2.2 \times 10^{14}$ GeV at 95\% credible intervals. 

In the last two decades, Bayesian statistics has  become the industry standard for parameter inference in almost all areas of Astrophysics and Cosmology~\cite{Trotta}, including in searches for LIV. However, there has been a renaissance in the use of frequentist statistics in the field of Cosmology, over the past 2-3 years, where the nuisance parameters were  dispensed  with using profile likelihood~\citep{Herold,Campeti,Colgain24,Karwal24,Herold24}. Some of the advantages and disadvantages of profile likelihood as compared  to Bayesian analysis have been reviewed in the aforementioned works.

%In a recent work, we have reanalyzed the spectral lag data for GRB 160625B for LIV using profile likelihood~\citep{Ganguly24}, to deal with astrophysical nuisance parameters, as a  complement to previous analyses of these data, which used Bayesian inference~\citep{Wei,Ganguly,Gunapati}. We showed that using profile likelihood,  we do not get a global minimum for $\chi^2$ as a function of $E_{QG}$ for all energies up to the Planck scale.
%This is in contrast to  Bayesian analysis, where we get  closed intervals for $E_{QG}$ well below the Planck scale~\citep{Wei,Gunapati}.  Therefore, for this GRB, the frequentist inference technique  allowed   us to set one-sided lower limits on $E_{QG}$ in a seamless manner. 

In this work, we redo the analysis in W17  using frequentist analysis, where we once again deal with nuisance parameters using profile likelihood.  This manuscript is structured as follows. The analysis methodology is described in Sec.~\ref{sec:analysis}. Our results are discussed in Sec.~\ref{sec:results}, and we summarize our conclusions in Sec.~\ref{sec:conclusions}. 

%%%%%%%%%%%%%%%%%%%%%%%%%%%%%%%%%%%%%%%%%%
\section{Analysis Methodology}
\label{sec:analysis}
We briefly recap the equations used for the analysis of LIV following the  same prescription and assumptions  as  in W17.  The observed spectral time lag ($\Delta t_{obs}$) from a given GRB at a redshift $z$ can be written down as
\begin{equation}
    \frac{\Delta t_{obs}}{1+z} = a_{LIV}K + \langle b \rangle,
    \label{eq:sum}
\end{equation}
where $a_{LIV} K$ is given by the following expression for superluminal LIV~\citep{Jacob}: 
\begin{equation}
    a_{LIV}K = \frac{1+n}{2H_0} \frac{E_h^n -E_l^n}{(1+z)^n E^n_{QG,n}} 
    \int_{0}^{z} \frac{(1+z^\prime)^n dz^\prime}{\sqrt{\Omega_M (1+z^\prime)^3 + 1-\Omega_M}},
    \label{eq:LIV}
\end{equation}
where $n$ indicated the order of LIV and is equal to 1 and 2 for linear and quadratic LIV, respectively; $\Omega_M$  and $H_0$ are the cosmological parameters corresponding to the matter density and Hubble constant, respectively.  We used the same values for the cosmological parameters as W17 (viz. $\Omega_M=0.308$ and $H_0=67.8$ km/sec/Mpc).
The energies $E_h$ and $E_l$  correspond to the energies in the rest frame bands, from which the spectral lags were obtained with $E_h>E_l$.
The second term in Eq.~\eqref{eq:sum}, namely  $\langle b \rangle$ represents the average effect of intrinsic time lags (due to astrophysics), as discussed in W17. Although a large number of phenomenological models have been used to model the intrinsic spectral lag~\citep{Desairev}, here we model the astrophysical  lag by a constant term similar to W17 for a straightforward comparison.

Similar to W17, we fit the observable $\frac{\Delta t_{obs}}{1+z}$ to Eq.~\eqref{eq:sum}  using maximum likelihood estimation and by adding an additional intrinsic scatter ($\sigma_{int}$) to the observed uncertainties in the spectral lags
\begin{equation}
     \mathcal{L}(E_{QG},\sigma_{int},\langle b \rangle) =\prod_{i=1}^N \frac{1}{\sqrt{\left(\frac{\sigma_{i}}{1+z}\right)^2+\sigma_{int}^2}} \exp \left\{-\frac{\left(\frac{\Delta t_{obs}}{1+z}-a_{LIV}K- \langle b \rangle \right)^2}{2\left(\sigma_{int}^2+\left(\frac{\sigma_{i}}{1+z}\right)^2\right) }\right\},
     \label{eq:likelihood}
  \end{equation}
where $\sigma_i$ is the uncertainty in $\Delta t_{obs}$, and $\sigma_{int}$ is the unknown intrinsic scatter, which we fit for.
Therefore, our regression problem contains three unknown parameters: $E_{QG}$, $\langle b \rangle$,  and $\sigma_{int}$
In this problem, $\langle b \rangle$ and $\sigma_{int}$ represent the nuisance parameters, which we account for using profile likelihood to get the likelihood distribution as a function of $E_{QG}$
\begin{equation}
\mathcal{L}(E_{QG})= \max_{\sigma_{int},\langle b \rangle} \mathcal{L}(E_{QG},\sigma_{int},\langle b \rangle)
\label{eq:PL}
\end{equation}
%Note that one difference compared to our previous work on profile likelihood~\cite{Ganguly24}, is that in this example, one free parameter ($\sigma_{int}$) appears both inside and outside the exponent in Eq.~\ref{eq:likelihood} unlike in Ref.~\cite{Ganguly24}.
For ease of computation, instead of maximizing Eq.~\eqref{eq:PL}
we construct    $\chi^2$ which is defined as
\begin{equation}
\chi^2 \equiv -2 \ln L (E_{QG},\sigma_{int},\langle b \rangle)
\label{eq:PLchi}
\end{equation}
We then minimize $\chi^2$ defined in Eq.~\eqref{eq:PLchi} over $\sigma_{int}$ and $\langle b \rangle$ for a fixed value of $E_{QG}$.  We then obtain frequentist confidence intervals (or upper limits) on $E_{QG}$ from $\Delta \chi^2 (E_{QG}) = \chi^2 (E_{QG}) - \chi^2_{min}$, where $\chi^2_{min}$ is the global minimum for $\chi^2$ over all values of $E_{QG}$. For this purpose, we use Wilks' theorem, which states that $\Delta\chi^2$ follows a $\chi^2$ distribution for one degree of freedom~\cite{Wilks1938}.

%%%%%%%%%%%%%%%%%%%%%%%%%%%%%%%%%%%%%%%%%%
\section{Results}
\label{sec:results}
We now apply the methodology in  the previous section to the  spectral lag data of 56 \rthis{Swift-BAT} detected GRBs , \rthis{consisting of both short and long GRBs collated} in ~\citet{Bernardini}, where the spectral lags have been calculated in fixed rest frame energy bands of 100-150 keV and 200-250 keV. \rthis{This dataset consists of GRBs with redshifts ranging from 0.35 (GRB 061021) to 5.47 (GRB 060927), having a mean redshift of 1.73. The energy gap between the midpoints of the successive rest-frame energy intervals is fixed at 100 keV.}
\rthis{The uncertainties in the spectral delay are calculated by averaging the left and right uncertainties provided in the aforementioned work. The full details of the 56 GRBs used for the analysis, such as the GRB name, redshift, observed spectral lags, and their uncertainties, can be found in Table 1 of ~\citet{Bernardini}.}

To evaluate the profile likelihood, we construct  a logarithmically spaced grid for $E_{QG}$ from $10^5$ GeV  to $10^{19}$ GeV for  linear and quadratic models of LIV.  The upper bound of $10^{19}$ GeV corresponds to the  Planck scale.  For each value of $E_{QG}$, we calculate the minimum value of $\chi^2 (E_{QG})$ by minimizing over $\sigma_{int}$ and $\langle b \rangle$. This minimization was done using the {\tt scipy.optimize.fmin} function, which uses the Nelder-Mead simplex algorithm~\citep{NR}. We also cross-checked this result with other minimization algorithms available in {\tt scipy} and found that the results do not change.

\begin{figure}[H]
    \centering
    \includegraphics[width=0.7\linewidth]{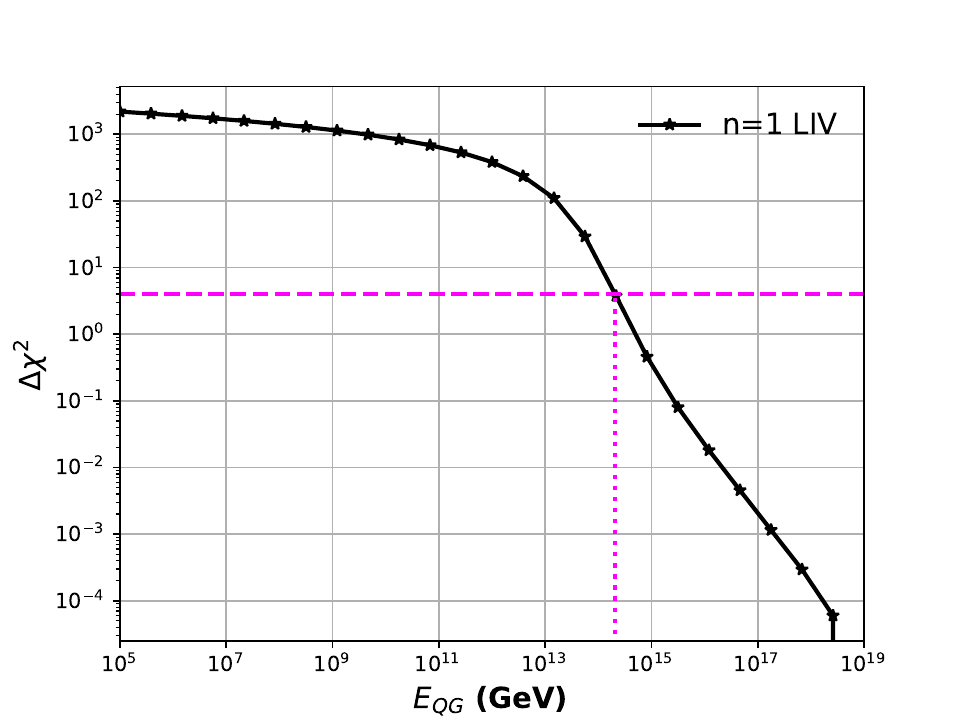}
    \caption{$\Delta\chi^2$, defined as ($\chi^2-\chi^2_{E_{pl}}$),  plotted against $E_{QG}$ for a linearly dependent LIV, corresponding to $n = 1$ , in Eq. \ref{eq:LIV}. The horizontal magenta dashed line represents $\Delta\chi^2 = 4$, and the vertical magenta dashed line provides us the x-intercept, the 95\% confidence level lower limit for $E_{QG} = 2.07\times 10^{14}$ GeV.}
    \label{fig:linearLIV}
\end{figure}

\begin{figure}[H]
    \centering
    \includegraphics[width=0.7\linewidth]{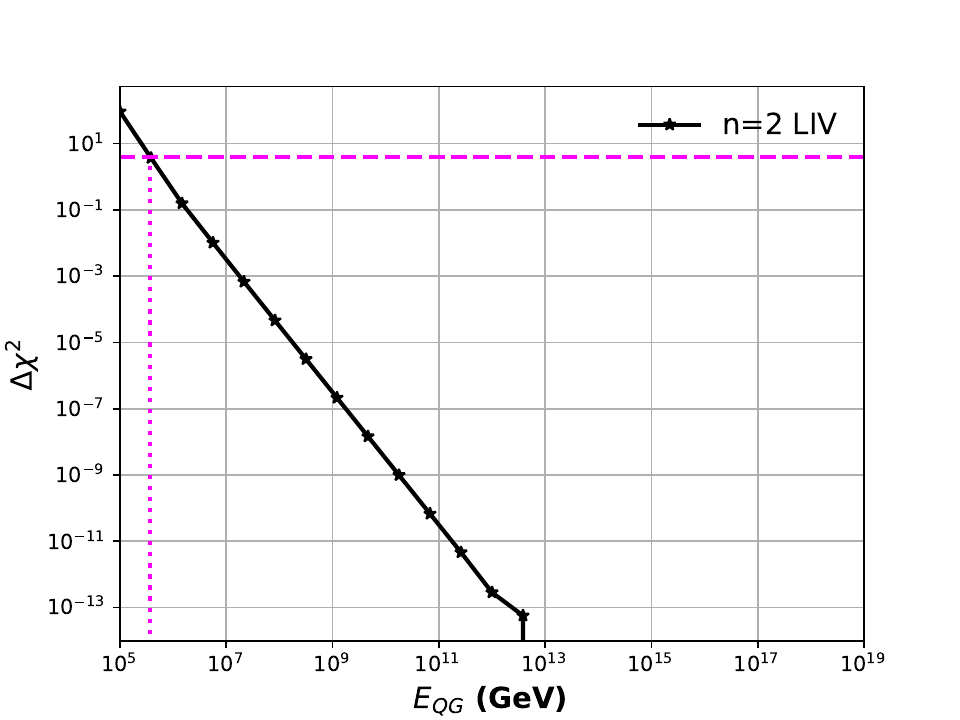}
    \caption{$\Delta\chi^2$, defined as ($\chi^2-\chi^2_{E_{pl}}$),  plotted against $E_{QG}$ for a quadratically dependent LIV, corresponding to $n = 2$, in Eq. \ref{eq:LIV}. The horizontal magenta dashed line represents $\Delta\chi^2 = 4$, and the vertical magenta dashed line provides us the x-intercept, the 95\% confidence level lower limit for $E_{QG} = 3.71\times 10^{5}$ GeV.}
    \label{fig:quadLIV}
\end{figure}

We find that $\chi^2$ does not achieve global minima below the Planck scale ($E_{pl}$). We then plot the curves of $\Delta \chi^2$ as a function of $E_{QG}$ for the linear and quadratic models of LIV, where  $\Delta \chi^2= \chi^2 (E_{QG}) -\chi^2 (E_{pl})$.  These $\Delta \chi^2$ curves can found in Fig.~\ref{fig:linearLIV} and Fig.~\ref{fig:quadLIV}, for linear and quadratic models of LIV, respectively. Since we do not obtain a global minima below the Planck scale, we can set one-sided \rthis{95.4\%} confidence level (c.l.) lower limits, by finding the x-intercept for which $\Delta \chi^2=4$. 
\rthis{For brevity, we denote 95.4\% c.l. as 95\% c.l.}
These 95\% c.l. lower limits are given by $E_{QG} \geq 2.07 \times 10^{14}$ GeV and $E_{QG}\geq  3.71\times 10^{5}$ GeV, for linear and quadratic LIV, respectively. Therefore,  we can set lower limits on the energy scale of LIV in a seamless manner, since we do not get a global minimum.
\rthis{Since our main aim was to compare our results to W17, we have used the same cosmological parameters as those in W17. When we vary the cosmological parameters and use the latest values from PDG, viz. $H_0=67.4$ km/sec and $\Omega_m=0.315$, we do not find qualitative differences
in the shape of $\chi^2$ as a function of $E_{QG}$. The new 95\% lower limits on $E_{QG}$ change to $E_{QG} \geq 2.08 \times 10^{14}$ GeV and $E_{QG}\geq  3.71\times 10^{5}$ GeV, for linear and quadratic LIV, respectively. Therefore, the variation in the lower limit on $E_{QG}$ is negligible upon choosing the PDG cosmological parameters. }

In order to judge the efficacy of the fit, similar to W17, we calculate the $\chi^2_{fit}$ based on the residuals between the data and best-fit model
\begin{equation}
\chi^2_{fit} =\sum_i \frac{\left(\frac{\Delta t_{i}}{1+z_{i}}-a_{\rm LIV}K_{i}-\langle b \rangle\right)^{2}}{\sigma_{\rm int}^{2}+\left(\frac{\sigma_i}{1+z_{i}\
}\right)^{2}}\;.
\label{eq:chifit}
\end{equation}
Note that $\chi^2_{fit}$ is used to ascertain the quality of the fit and is 
different from $\chi^2$ defined in Eq.~\eqref{eq:PLchi}. We evaluated $\chi^2_{fit}$ and $\chi^2_{fit}$/DOF for values of $E_{QG}$ at  both the Planck scale and the 95\% cl. lower limit on $E_{QG}$ for both the LIV models. Here, DOF refers to the degrees of freedom, which is equal to the difference between  total number of data points and number of free parameters (three).
These  values can be found in Table~\ref{tab:model_comp}. We also show the best-fit values of $\langle b \rangle$ and $\sigma_{int}$ in this table. We find that the best-fit values of $\langle b \rangle$ and $\sigma_{int}$ are consistent within the 68\% credible regions for the  marginalized posteriors obtained  in W17 (for linear LIV). The reduced $\chi^2_{fit}$ is close to one for both models, although it includes an intrinsic scatter of about 2\%.

\begin{table}[h]
    \centering
    \renewcommand{\arraystretch}{1.5}
    \begin{tabular}{|c|cc|cc|}
         \hline
         {} & \multicolumn{2}{c|}{\textbf{Linear LIV}} & \multicolumn{2}{c|}{\textbf{Quadratic LIV}}\\
         {} & \multicolumn{2}{c|}{$\mathbf{n=1}$} & \multicolumn{2}{c|}{$\mathbf{n=2}$}\\
         \hline
         $\langle b \rangle$ & \multicolumn{2}{c|}{$-0.035$} & \multicolumn{2}{c|}{$-0.011$}\\
         $\sigma_{int}$ & \multicolumn{2}{c|}{0.023} & \multicolumn{2}{c|}{0.023}\\
         $E_{QG}$ (GeV) & \multicolumn{2}{c|}{$2.07 \times 10^{14}$} & \multicolumn{2}{c|}{$3.71 \times 10^{5}$}\\
         \hline
         {} & $\mathbf{E_{QG}}$ & $\mathbf{E_{pl}}$ & $\mathbf{E_{QG}}$ & $\mathbf{E_{pl}}$\\
         {$\chi^2_{fit}$}/DOF & $52.88/54$ & $53.91/54$ & $52.46/54$ & $53.95/54$\\
         \hline
    \end{tabular}
    \caption{Best-fit model parameters: $\langle b \rangle$ and $\sigma_{int}$, corresponding to both linear and quadratic LIV models, evaluated at their respective 95\% confidence lower limits of $E_{QG}$. The standard frequentist goodness-of-fit metric, $\chi^2_{fit}/DOF$ (refer Eq.\ref{eq:chifit}), is also reported at both the 95\% c.l. lower limit for $E_{QG}$  and at the Planck scale. Here, DOF refers to the degrees of freedom, which is equal to the difference between  total number of data points and number of free parameters.}
    \label{tab:model_comp}
\end{table}

%%%%%%%%%%%%%%%%%%%%%%%%%%%%%%%%%%%%%%%%%%
\section{Conclusions}
\label{sec:conclusions}
In this work, we have reanalyzed the data for spectral lags of 56 GRBs between two fixed energy bands in the rest frame, in order to search for LIV  using frequentist inference. For this analysis, we use profile likelihood to deal with the astrophysical nuisance parameters, and set a constraint on the energy scale of LIV for both linear and quadratic  models.

We parametrize  the rest frame spectral lags as a sum of a constant intrinsic lag and LIV induced time lag. Similarly to W17, we use a Gaussian likelihood and also incorporate another free parameter for the intrinsic scatter, which is added in quadrature to the observed uncertainties in the spectral lags. Therefore, our regression model consists of two nuisance parameters and one physically interesting parameter, viz. the energy scale for LIV.

We find that after dealing with nuisance parameters using profile likelihood, we do not find a global minimum for $\chi^2$ as a function of $E_{QG}$ below the Planck energy scale. These plots for $\Delta \chi^2$ as a function of LIV for both the linear and quadratic models of LIV are shown in Fig.~\ref{fig:linearLIV} and Fig.~\ref{fig:quadLIV}, respectively. Therefore, we can set one-sided lower limits at any confidence levels from the X-intercept of the $\Delta \chi^2$ curves.
These 95\% confidence level lower limits obtained from $\Delta \chi^2=4$ are given 
by $E_{QG}\geq 2.07 \times 10^{14}$ GeV and $E_{QG}\geq  3.71\times 10^{5}$ GeV, for linear and quadratic LIV, respectively. Our lower limit for linear LIV is comparable to the value obtained in W17 ($2.2 \times 10^{14}$ GeV).
The best-fit values for the two nuisance parameters evaluated at  two different energies (Planck scale and 95\% c.l. lower limit value) can be found in Table~\ref{tab:model_comp}.

Therefore, we have shown that the profile likelihood method provides a viable alternative in dealing with nuisance parameters, which is complementary to the widely used Bayesian inference technique, and for our example, allows us to seamlessly infer the lower limits in an automated manner.  In the spirit of open science, we have made our analysis codes publicly available, which can be found on \href{https://github.com/vyaas3305/liv-PL-restframe}{Github}

%%%%%%%%%%%%%%%%%%%%%%%%%%%%%%%%%%%%%%%%%%
\vspace{6pt} 

\begin{adjustwidth}{-\extralength}{0cm}

\reftitle{References}


% \bibliography{your_external_BibTeX_file}

%=====================================
% References, variant B: internal bibliography
%=====================================
% APA format (Used for journal: admsci, behavsci, businesses, econometrics, economies, education, ejihpe, games, humans, ijfs, journalmedia, jrfm, languages, psycholint, publications, tourismhosp, youth)
%\isAPAStyle{%
%\begin{thebibliography}{999}
\bibliography{main}

\begin{thebibliography}{999}

\bibitem[{Desai}(2024)]{Desairev}
{Desai}, S.
\newblock {Astrophysical and Cosmological Searches for Lorentz Invariance Violation}. In {\em Recent Progress on Gravity Tests. Challenges and Future Perspectives}; {Bambi}, C.; {C{\'a}rdenas-Avenda{\~n}o}, A., Eds.;  2024; pp. 433--463.
\newblock {\url{https://doi.org/10.1007/978-981-97-2871-8_11}}.

\bibitem[{Yu} et~al.(2022){Yu}, {Gao}, {Wang}, and {Zhang}]{WuGRBreview}
{Yu}, Y.W.; {Gao}, H.; {Wang}, F.Y.; {Zhang}, B.B.
\newblock {Gamma-Ray Bursts}. In {\em Handbook of X-ray and Gamma-ray Astrophysics. Edited by Cosimo Bambi and Andrea Santangelo};  2022; p.~31.
\newblock {\url{https://doi.org/10.1007/978-981-16-4544-0_126-1}}.

\bibitem[{Wei} and {Wu}(2022)]{WeiWu2}
{Wei}, J.J.; {Wu}, X.F.
\newblock {Tests of Lorentz Invariance}. In {\em Handbook of X-ray and Gamma-ray Astrophysics. Edited by Cosimo Bambi and Andrea Santangelo};  2022; p.~82.
\newblock {\url{https://doi.org/10.1007/978-981-16-4544-0_132-1}}.

\bibitem[{Wei} and {Wu}(2017)]{WeiWu}
{Wei}, J.J.; {Wu}, X.F.
\newblock {A Further Test of Lorentz Violation from the Rest-frame Spectral Lags of Gamma-Ray Bursts}.
\newblock {\em \apj} {\bf 2017}, {\em 851},~127,  \href{http://arxiv.org/abs/1711.09185}{{\normalfont [arXiv:astro-ph.HE/1711.09185]}}.
\newblock {\url{https://doi.org/10.3847/1538-4357/aa9d8d}}.

\bibitem[{Bernardini} et~al.(2015){Bernardini}, {Ghirlanda}, {Campana}, {Covino}, {Salvaterra}, {Atteia}, {Burlon}, {Calderone}, {D'Avanzo}, {D'Elia}, {Ghisellini}, {Heussaff}, {Lazzati}, {Melandri}, {Nava}, {Vergani}, and {Tagliaferri}]{Bernardini}
{Bernardini}, M.G.; {Ghirlanda}, G.; {Campana}, S.; {Covino}, S.; {Salvaterra}, R.; {Atteia}, J.L.; {Burlon}, D.; {Calderone}, G.; {D'Avanzo}, P.; {D'Elia}, V.;  et~al.
\newblock {Comparing the spectral lag of short and long gamma-ray bursts and its relation with the luminosity}.
\newblock {\em \mnras} {\bf 2015}, {\em 446},~1129--1138,  \href{http://arxiv.org/abs/1410.5216}{{\normalfont [arXiv:astro-ph.HE/1410.5216]}}.
\newblock {\url{https://doi.org/10.1093/mnras/stu2153}}.

\bibitem[{Trotta}(2017)]{Trotta}
{Trotta}, R.
\newblock {Bayesian Methods in Cosmology}.
\newblock {\em ArXiv e-prints} {\bf 2017},  \href{http://arxiv.org/abs/1701.01467}{{\normalfont [1701.01467]}}.

\bibitem[{Herold} et~al.(2022){Herold}, {Ferreira}, and {Komatsu}]{Herold}
{Herold}, L.; {Ferreira}, E.G.M.; {Komatsu}, E.
\newblock {New Constraint on Early Dark Energy from Planck and BOSS Data Using the Profile Likelihood}.
\newblock {\em \apjl} {\bf 2022}, {\em 929},~L16,  \href{http://arxiv.org/abs/2112.12140}{{\normalfont [arXiv:astro-ph.CO/2112.12140]}}.
\newblock {\url{https://doi.org/10.3847/2041-8213/ac63a3}}.

\bibitem[{Campeti} and {Komatsu}(2022)]{Campeti}
{Campeti}, P.; {Komatsu}, E.
\newblock {New Constraint on the Tensor-to-scalar Ratio from the Planck and BICEP/Keck Array Data Using the Profile Likelihood}.
\newblock {\em \apj} {\bf 2022}, {\em 941},~110,  \href{http://arxiv.org/abs/2205.05617}{{\normalfont [arXiv:astro-ph.CO/2205.05617]}}.
\newblock {\url{https://doi.org/10.3847/1538-4357/ac9ea3}}.

\bibitem[{Colg{\'a}in} et~al.(2024){Colg{\'a}in}, {Pourojaghi}, and {Sheikh-Jabbari}]{Colgain24}
{Colg{\'a}in}, E.{\'O}.; {Pourojaghi}, S.; {Sheikh-Jabbari}, M.M.
\newblock {Implications of DES 5YR SNe Dataset for $\Lambda$CDM}.
\newblock {\em arXiv e-prints} {\bf 2024}, p. arXiv:2406.06389,  \href{http://arxiv.org/abs/2406.06389}{{\normalfont [arXiv:astro-ph.CO/2406.06389]}}.
\newblock {\url{https://doi.org/10.48550/arXiv.2406.06389}}.

\bibitem[{Karwal} et~al.(2024){Karwal}, {Patel}, {Bartlett}, {Poulin}, {Smith}, and {Pfeffer}]{Karwal24}
{Karwal}, T.; {Patel}, Y.; {Bartlett}, A.; {Poulin}, V.; {Smith}, T.L.; {Pfeffer}, D.N.
\newblock {Procoli: Profiles of cosmological likelihoods}.
\newblock {\em arXiv e-prints} {\bf 2024}, p. arXiv:2401.14225,  \href{http://arxiv.org/abs/2401.14225}{{\normalfont [arXiv:astro-ph.CO/2401.14225]}}.
\newblock {\url{https://doi.org/10.48550/arXiv.2401.14225}}.

\bibitem[{Herold} et~al.(2025){Herold}, {Ferreira}, and {Heinrich}]{Herold24}
{Herold}, L.; {Ferreira}, E.G.M.; {Heinrich}, L.
\newblock {Profile likelihoods in cosmology: When, why, and how illustrated with <inline-formula><mml:math><mml:mi>{\ensuremath{\Lambda}}</mml:mi><mml:mi>CDM</mml:mi></mml:math></inline-formula>, massive neutrinos, and dark energy}.
\newblock {\em \prd} {\bf 2025}, {\em 111},~083504,  \href{http://arxiv.org/abs/2408.07700}{{\normalfont [arXiv:astro-ph.CO/2408.07700]}}.
\newblock {\url{https://doi.org/10.1103/PhysRevD.111.083504}}.

\bibitem[{Jacob} and {Piran}(2008)]{Jacob}
{Jacob}, U.; {Piran}, T.
\newblock {Lorentz-violation-induced arrival delays of cosmological particles}.
\newblock {\em \jcap} {\bf 2008}, {\em 1},~031,  \href{http://arxiv.org/abs/0712.2170}{{\normalfont [0712.2170]}}.
\newblock {\url{https://doi.org/10.1088/1475-7516/2008/01/031}}.

\bibitem[Wilks(1938)]{Wilks1938}
Wilks, S.S.
\newblock The large-sample distribution of the likelihood ratio for testing composite hypotheses.
\newblock {\em The annals of mathematical statistics} {\bf 1938}, {\em 9},~60--62.

\bibitem[{Press} et~al.(1992){Press}, {Teukolsky}, {Vetterling}, and {Flannery}]{NR}
{Press}, W.H.; {Teukolsky}, S.A.; {Vetterling}, W.T.; {Flannery}, B.P.
\newblock {\em {Numerical recipes in FORTRAN. The art of scientific computing}};  1992.

\end{thebibliography}
%\end{thebibliography}
%}{}

% If authors have biography, please use the format below
%\section*{Short Biography of Authors}
%\bio
%{\raisebox{-0.35cm}{\includegraphics[width=3.5cm,height=5.3cm,clip,keepaspectratio]{Definitions/author1.pdf}}}
%{\textbf{Firstname Lastname} Biography of first author}
%
%\bio
%{\raisebox{-0.35cm}{\includegraphics[width=3.5cm,height=5.3cm,clip,keepaspectratio]{Definitions/author2.jpg}}}
%{\textbf{Firstname Lastname} Biography of second author}

% For the MDPI journals use author-date citation, please follow the formatting guidelines on http://www.mdpi.com/authors/references
% To cite two works by the same author: \citeauthor{ref-journal-1a} (\citeyear{ref-journal-1a}, \citeyear{ref-journal-1b}). This produces: Whittaker (1967, 1975)
% To cite two works by the same author with specific pages: \citeauthor{ref-journal-3a} (\citeyear{ref-journal-3a}, p. 328; \citeyear{ref-journal-3b}, p.475). This produces: Wong (1999, p. 328; 2000, p. 475)

%%%%%%%%%%%%%%%%%%%%%%%%%%%%%%%%%%%%%%%%%%
%% for journal Sci
%\reviewreports{\\
%Reviewer 1 comments and authors’ response\\
%Reviewer 2 comments and authors’ response\\
%Reviewer 3 comments and authors’ response
%}
%%%%%%%%%%%%%%%%%%%%%%%%%%%%%%%%%%%%%%%%%%
\section*{Appendix}
\rthis{In order to ensure that the differences in our results between frequentist and Bayesian analysis are not due to statistical fluctuations, we redo the analysis by doing a random swapping of parameter uncertainties among the 56 GRBs, in order to check the robustness of our results. 
To avoid altering the redshifts and the spectral lags of the observed GRBs, we only swap the uncertainties in the spectral lags amongst the GRBs, and then repeat both Bayesian as well as frequentist analysis. All the other parameters therefore remain the same. For Bayesian analysis, we use the same likelihood as in Eq.~\eqref{eq:likelihood} and  uniform priors on $\log (E_{QG})$, $b$, and $\sigma_{int}$, given by $\mathcal{U}$ (1,19), $\mathcal{U}$ (-1.0,1.0), and 
$\mathcal{U}$ (0,1.0), respectively. }

\rthis{The results of the profile likelihood analysis can be found in Fig.~\ref{fig:linearLIV_shuff} and Fig.~\ref{fig:quadLIV_shuff} for linear and quadratic LIV, respectively. We find that $\Delta \chi^2$ plot as a function of $E_{QG}$   does not show a global minimum for the linear model of LIV, similar to before. The 95\% c.l. lower limit on $E_{QG}$ using these shuffled uncertainties is given by $E_{QG} \geq 3.1\times 10^{14}$ GeV. For the quadratic LIV case, although we get a global minimum below the Planck scale, we find that $\Delta \chi^2$ asymptotes towards a constant value of 1.4 above the global minima. Therefore, similar to before, we can only set one-sided lower limits on $E_{QG}$ at 95\% c.l., since $\Delta \chi^2$ does not exceed the value of 4.0 after its minimum.
Therefore, the lower limit on  $E_{QG}$ at 95\% c.l. is given by  $E_{QG} \geq 1.8\times 10^{5}$ GeV. Therefore, these frequentist limits limits using the shuffled data are of the same order of magnitude as that obtained using the original data.}

\rthis{The corresponding Bayesian credible intervals for all the three free parameters can be found in Fig.~\ref{fig:linearLIV_bayesian} and Fig.~\ref{fig:quadLIV_bayesian}, for linear and quadratic LIV models, respectively. We find that we don't get bounded  marginalized 95\% credible intervals for $E_{QG}$ for both linear as well as quadratic LIV models. This behavior is qualitatively similar to that seen for frequentist analysis. Therefore, we can only set one-sided Bayesian lower limits. These lower limits at 95\% credible intervals are given by $E_{QG} \geq 3.1\times 10^{14}$  GeV and $E_{QG} \geq 1.8\times 10^{5}$ GeV for linear and quadratic LIV, respectively. These limits marginally differ from the corresponding frequentist lower limits, although are of the same order of magnitude.}

\rthis{Therefore, we find that once we shuffle  the uncertainties in the spectral lags, although the  we don't get closed intervals for $E_{QG}$ at 95\% confidence/credible intervals using both frequentist and Bayesian analysis, the limits are marginally different. However, we note these results are using only one  realization of the bootstrapped data. } 

 \begin{figure}[H]
    \centering
    \includegraphics[width=0.7\linewidth]{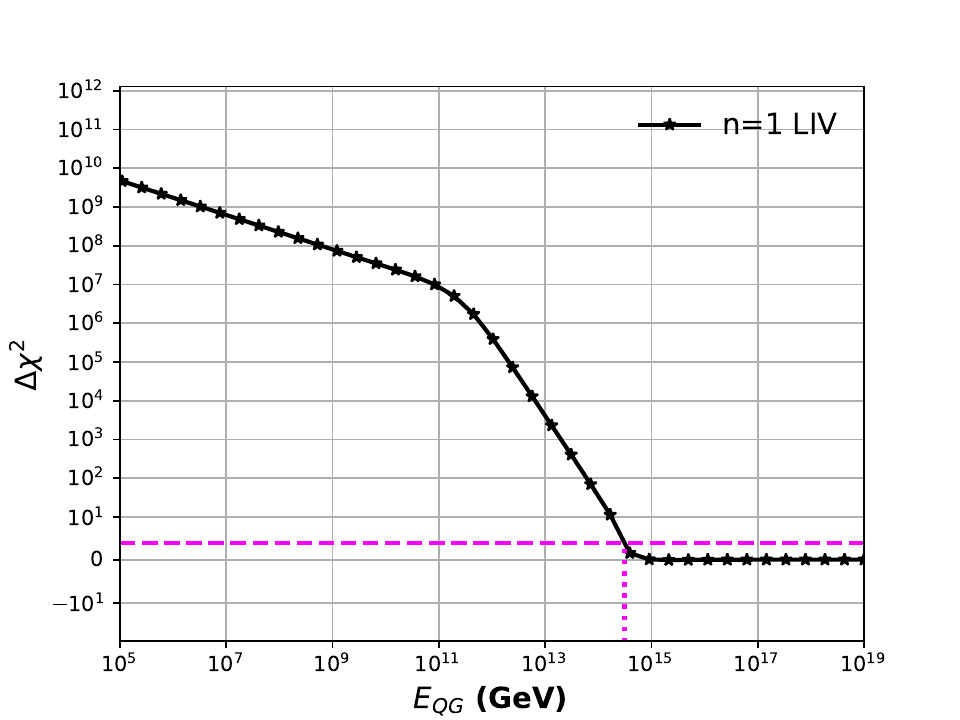}
    \caption{$\Delta\chi^2$, defined as ($\chi^2-\chi^2_{E_{pl}}$),  plotted against $E_{QG}$ for a linearly dependent LIV, corresponding to $n = 1$ , in Eq. \eqref{eq:LIV}, after shuffling the uncertainties in the spectral lags among the GRBs. The horizontal magenta dashed line represents $\Delta\chi^2 = 4$, and the vertical magenta dashed line provides us the x-intercept, the 95\% confidence level lower limit for $E_{QG} = 3.1\times 10^{14}$ GeV.}
    \label{fig:linearLIV_shuff}
\end{figure}

\begin{figure}[H]
    \centering
    \includegraphics[width=0.7\linewidth]{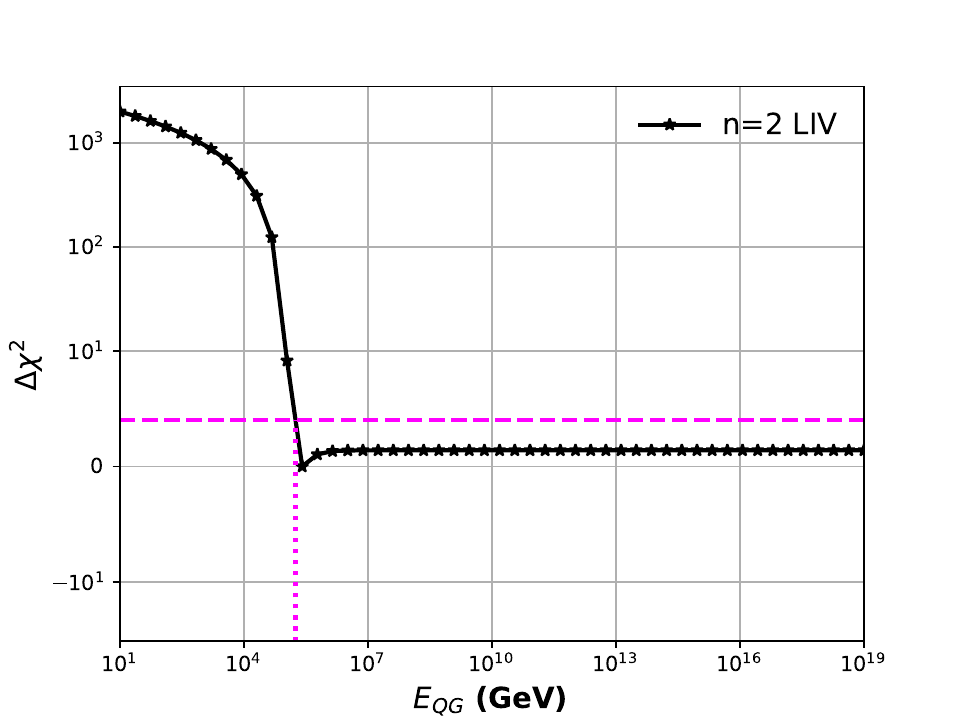}
    \caption{$\Delta\chi^2$, defined as ($\chi^2-\chi^2_{E_{pl}}$),  plotted against $E_{QG}$ for a quadratically dependent LIV, corresponding to $n = 2$, in Eq. \eqref{eq:LIV}, after shuffling the uncertainties in the spectral lags among the GRBs. The horizontal magenta dashed line represents $\Delta\chi^2 = 4$, and the vertical magenta dashed line provides us the x-intercept, the 95\% confidence level lower limit for $E_{QG} = 1.8\times 10^{5}$ GeV.}
    \label{fig:quadLIV_shuff}
\end{figure}

\begin{figure}[H]
    \centering
    \includegraphics[width=0.7\linewidth]{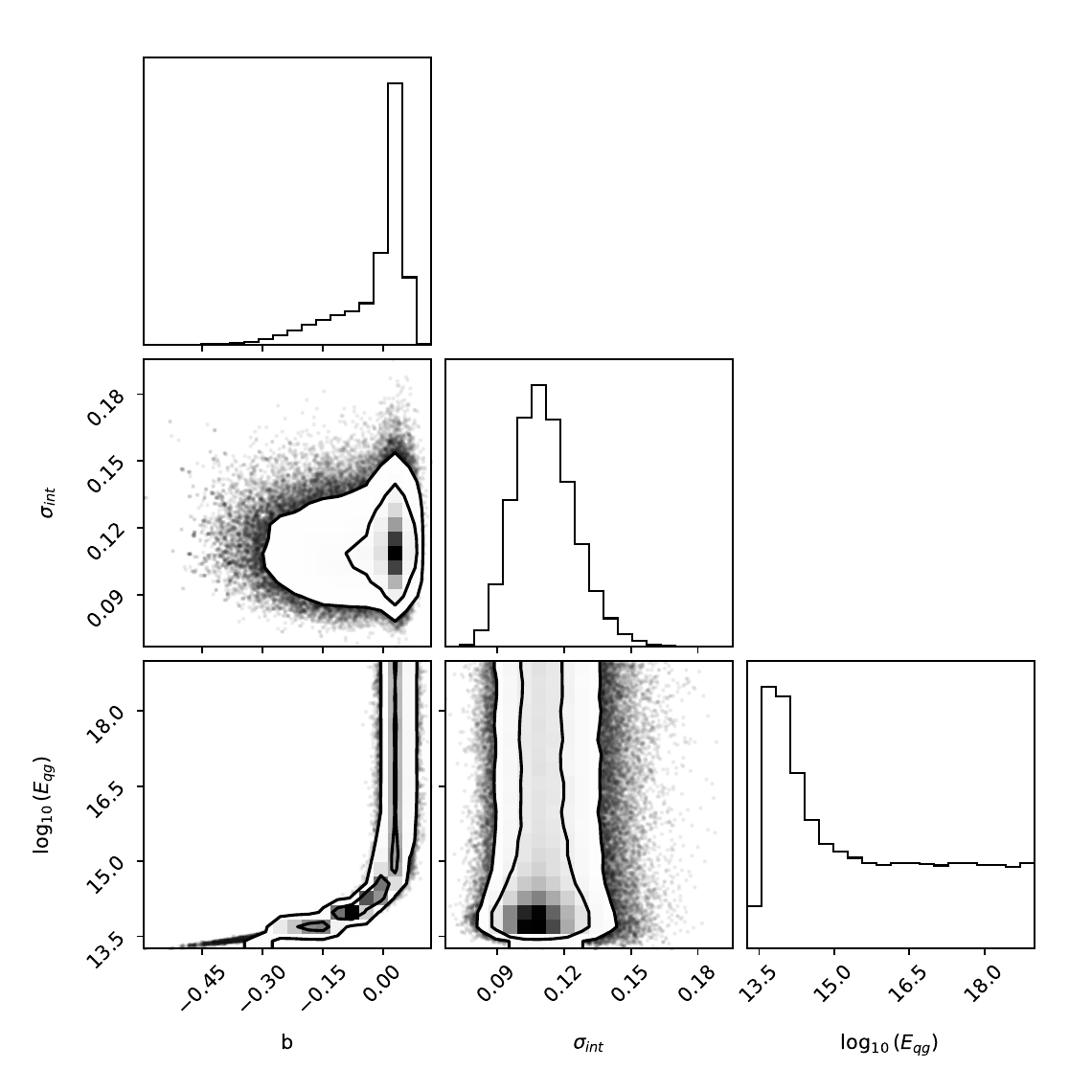}
    \caption{The marginalized contours for $E_{QG}$, $b$ and $\sigma_{int}$ at 68\% and 95\% credible intervals for  linear model of LIV, corresponding to $n = 1$, in Eq.~\eqref{eq:LIV}. The corresponding 95\% lower limit for $E_{QG}$ is given by $E_{QG} = 3.66\times 10^{13}$ GeV.}
    \label{fig:linearLIV_bayesian}
\end{figure}

\begin{figure}[H]
    \centering
    \includegraphics[width=0.7\linewidth]{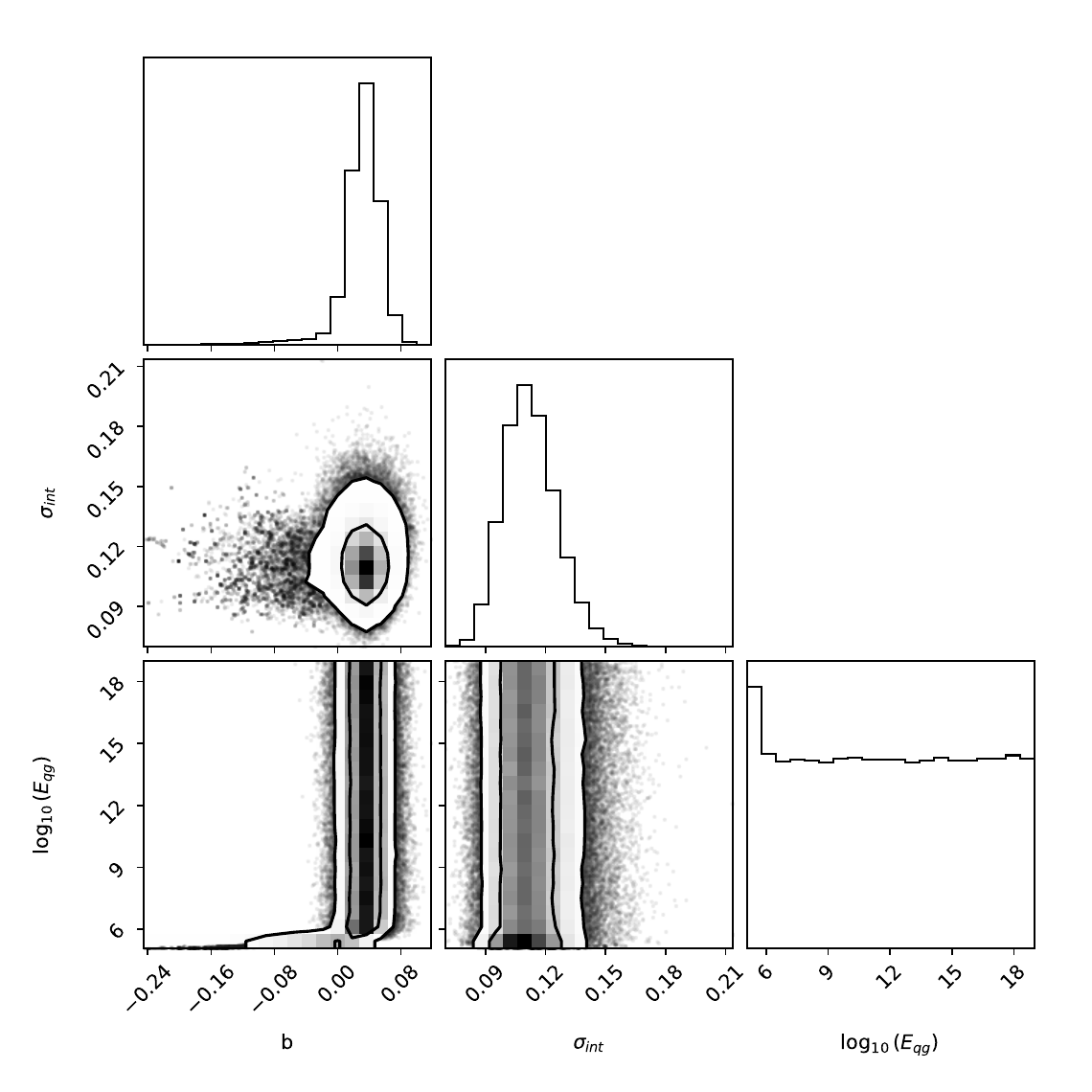}
    \caption{The marginalized contours for $E_{QG}$, $b$ and $\sigma_{int}$ at 68\% and 95\% credible intervals for quadratic  model of LIV, corresponding to $n = 2$, in Eq.~\eqref{eq:LIV} The corresponding 95\% lower limit for $E_{QG}$ is given by $E_{QG} = 2.01\times 10^{5}$ GeV.}
    \label{fig:quadLIV_bayesian}
\end{figure}

\PublishersNote{}
%\isPreprints{}{% This command is only used for ``preprints''.
\end{adjustwidth}
\end{document}